\documentclass[a4paper,12pt]{article}
\usepackage{graphicx}
\usepackage{amssymb}
\usepackage{amsmath}
\usepackage[a4paper,portrait,twocolumn=false,includeheadfoot,mag=1000,dvips]{geometry}
\usepackage[title]{appendix}
\usepackage{caption}
\usepackage{float}
\usepackage{color}
\usepackage{ulem}
\usepackage{mathrsfs}
\usepackage{type1cm}
\usepackage{marvosym}
\usepackage{amssymb}
\usepackage{wasysym}
\usepackage{subfigure}
\usepackage{multirow}
\usepackage{booktabs}
\usepackage[utf8]{inputenc}

\usepackage{graphicx}
\usepackage{epstopdf}

\usepackage{makecell}

\usepackage{enumerate}

\makeatletter

\newcommand{\Rmnum}[1]{\expandafter\@slowromancap\romannumeral #1@}
\makeatother

\numberwithin{equation}{section}

\newcommand{\xiaosihao}{\fontsize{14pt}{\baselineskip}\selectfont}

\begin{document}  \xiaosihao

\newtheorem{The}{Theorem}
\newtheorem{exam}{Example}
\newtheorem{lem}{Lemma}
\newtheorem{de}{Definition}
\newtheorem{prop}{Proposition}
\newtheorem{cor}{Corollary}
\newtheorem{hyp}{Hypothesis}
\newtheorem{rem}{Remark}
\pagestyle{plain}

\setlength{\baselineskip}{20pt}
\setlength{\parskip}{0.4\baselineskip}

\clearpage
\begin{center}
{\xiaosihao \textbf{Bernoulli Trials With Skewed Propensities } \\
\textbf{for}  \\ \textbf{Certification and Validation}
}

{\normalsize by}

{\normalsize Nozer D. Singpurwalla \\ The George Washington University, Washington, D.C. \\ and \\ Boya Lai \\}
{\normalsize The City University of Hong Kong, Hong Kong \\}
{\normalsize January 2020}


%

\end{center}

%
\renewcommand{\abstractname}{{\large Abstract}}
\begin{abstract}
{\normalsize
The impetus for writing this paper are the well publicized media reports that software failure was the cause of the two recent mishaps of the Boeing 737 Max aircraft. The problem considered here though, is a specific one, in the sense that it endeavors to address the general matter of conditions under which an item such as a drug, a material specimen, or a complex, system can be \textbf{\textit{certified}} for use based on a large number of Bernoulli trials, all successful. More broadly, the paper is an attempt to answer the old and honorable philosophical question, namely,``when can empirical testing on its own \textbf{\textit{validate}} a law of nature?'' Our message is that the answer depends on what one starts with, namely, what is one's prior distribution, what unknown does this prior distribution endow, and what has been observed as data.\\ [-15pt]

The paper is expository in that it begins with a historical overview, and ends with some new ideas and proposals for addressing the question posed. In the sequel, it also articulates on Popper's notion of ``\textbf{\textit{propensity}}'' and its role in providing a proper framework for Bayesian inference under Bernoulli trials, as well as the need to engage with posterior distributions that are \textit{subjectively specified}; that is, without a recourse to the usual Bayesian prior to posterior iteration.}\\ [10pt]
{\normalsize \textbf{Keywords:} \textit{Bayes-Laplace Priors, Jeffreys' Prior, Probabilistic Induction, Exchangeability, Drug Approval, Subjective Posteriors, Machine Learning}.\\}
\end{abstract}
\addtocounter{section}{-1}
\section{\large Preamble: Perspective}

Despite the science and engineering oriented implication of the title of this paper, its essential contribution is to address the general philosophical question as to whether empirical evidence on its own can ever prove a law of nature. The related practical question is when can a complex system, like computer software, a new drug, or an autonomous system, be certified for use based on testing alone. Karl Pearson (1920) characterized this matter as a ``Fundamental Problem of Practical Statistics''; his foresight continues to hold almost 100 years later. Later on, in (1926), E.B. Wilson wrote an article alluding to this matter which the June 2015 issue of the Proceedings of the  (US) National Academy of Sciences has labelled ``\textit{ A Sleeping Beauty in Science''}.

Whereas the problem considered is as ancient as Hume, Newton, and Kant, it was Bayes in 1763, followed by Laplace in 1774, who proposed \textit{\textbf{probabilistic induction}}, as a definitive response. Indeed, this was one of their signal achievements. As a matter of perspective, Bayes' Law turns out to be merely a methodological by-product of the bigger philosophical, ``\textit{problem with induction}'', of Thomas Hobbs (1588-1679), that Bayes and Laplace were trying to address. In effect, probabilistic induction can be seen as a compromise between objections to the principles of induction and of deduction. An interested reader may also want to see Zabell (1989) for a more philosophical as well as a detailed technical discourse on the problem.

With the advent of Data Science and Machine Learning, one is tempted to re-visit the broader question and ask, ``Can Big Data and Machine Learning, on their own, certify a complex system, or ever prove a Law of Nature?'' Were we to adopt a Bayes-Laplace like disposition, then we would arrive upon the viewpoint that data science and machine learning are the modern day \underline{enablers} of probabilistic induction. Computer scientists may disagree with this viewpoint, and declare it limited. All the same, the material of this paper can help address the nagging question as to whether, and when, can AI based systems, like self-driving cars, and other robotic entities be trusted for routine use.

\enlargethispage{\baselineskip}
The matter of certification and/or asserting a law of nature based on evidence alone can be explicated via the archetypal example of Bernoulli trials under \textit{\textbf{almost identical conditions}}. For the purpose of the next few sections, we take the notion ``almost identical'' as an \underline{undefined} \underline{primitive}, and refrain from any attempt to make it more precise.

\section{\large The Bernoulli Trials Framework}

Consider an experiment whose purpose is to certify a product, or to ascertain a law of nature, which entails $n$ Bernoulli trials, where the outcome of each trial is a success, if the law be true, and failure otherwise; similarly with the product. Suppose that $N$ such Bernoulli trials are to be performed under similar conditions, and let $R$ be the number of trials leading to a success; $0\leq R\leq N$. To assert a law, or to certify a product, we need to think of $N$ as being very large, conceptually infinite. The law is deemed asserted if $R=N$, but since N is infinite, we are unable to conduct all $N$ trials; thus $R$ will be unknown. To overcome this obstacle, $n$ out of the $N$ trials are chosen at random (i.e. without prejudice) and tested for successes and failures. Suppose that all the $n$ trials result in success. Based on the above, can one declare with certitude, that $R=N$, even if $n$ is very, very large?

Because of well documented difficulties with induction, the answer is an emphatic no! Indeed, from a mathematical point of view, one cannot invoke the inductive hypothesis, because $n$ out of $n$ successes does not imply $(n+1)$ out of $(n+1)$ successes. However, under probabilistic induction, one is able to assert that, under certain conditions, with $n$ out of $n$ successes at hand, and $n$ large, $P(R=N)\approx1$; $P$ denotes probability. In this paper, all probabilities are personal. Strategies for articulating these conditions, and approaches for assessing $P(R=N)$, have been the topic of many investigations; some are highlighted, in Section 2.

Our focus is on Bayesian approaches; that is methods which assign priors and invoke Bayes' formula. There are two general directions in which this has been done. The first is the classical approach of Bayes and Laplace in which a prior probability is assigned to the various values that $R$ can take. The second is, by now, a commonly used modern approach, whose foundation lies in De Finetti's famous representation theorem on \textit{\textbf{exchangeable sequences}}.

\section{\large The Bayes-Laplace Approach and Its Variants}

Suppose, as a start, that $N$ is finite, so that the unknown $R$ can take values, $0,1,\cdots,N$. Then, in a sample of size $n$, the probability of observing $T$ successes, is given by the hypergeometric distribution, so that
\begin{eqnarray*}
P(T=n|R)&=& \frac{
  \left(
  \begin{array}{c}
  R \\ n
  \end{array}
  \right)
}{
  \left(
  \begin{array}{c}
  N \\ n
  \end{array}
  \right)} ,  \quad R=n, n+1, \cdots, N \\
     &=& 0 ,  \quad \quad \quad \quad R=0, 1, \cdots, N-1.
\end{eqnarray*}
Our aim is to find $P(R=N|T=n)$ which by an application of Bayes' Law, with $P(R=r)$ as a prior for $R$, is
\begin{eqnarray*}
P(R=N|T=n) = \frac{
  P(R=N)
}{
  \sum\limits_{r=n}^{N}\frac{
  \left(
  \begin{array}{c}
  r \\ n
  \end{array}
  \right)}{\left(
  \begin{array}{c}
  N \\ n
  \end{array}
  \right)}P(R=r)
  }
\end{eqnarray*}
Thus all that is required to assess the probability that $(R=N)$ is a prior distribution for R. The essential spirit of the Bayes-Laplace approach (and its variants) is the assignment of a prior distribution on observables, like $R$. Of the several priors on R that have been considered, those given in Sections 2.1-2.4 are noteworthy. Also discussed therein are the consequences of each prior vis a vis their relevance to certification and validation.

\subsection{The Bayes-Laplace Prior}

The Bayes-Laplace strategy is that of expressing prior indifference among all possible values that $R$ can take, namely, $R=0,1,\cdots,N$. This is tantamount to a discrete uniform prior on $R$, namely $P(R=r)=(N+1)^{-1}$, for $r=0,1,\cdots,N$. Such a prior accords well with the principle of insufficient reason, and is therefore also known as a \textit{\textbf{public prior}}.

Under this prior, it can be seen that $P(R=N|T=n)=(n+1)/(N+1)$, so that for any fixed and large $n$, $\lim\limits_{N\rightarrow \infty} (n+1)/(N+1)=0$. This means that even if all the $n$ observed trials result in success, then irrespective of how large $n$ is, the probability that all future $N$ trials, (where $N$ is large), will lead to success is zero! A result such as this goes against the grain of experimental scientists, who having observed a slew of successes would balk at the prospect of being told that the probability of all future trials being a success is zero; they would prefer that the answer be something closer to one, if not one itself.

\subsection{Jeffrey's Prior}

As a reaction to the above concern spawned by the Bayes-Laplace prior, Jeffreys (1961) proposed a prior which places a large point mass at 0 and at N, and spreads the difference over the remaining values of $R$. Specifically
\begin{eqnarray*}
  P(R=N)= \bigg \{
  \begin{array}{l}
  \frac{1-2k}{N+1},  \quad \quad \ \ r=1,\cdots,N-1 \\
  \frac{1-2k}{N+1}+k, \quad r=0, N,
  \end{array}
\end{eqnarray*}
for $0<k\leq \frac{1}{2}$. For $k=0$, Jeffreys prior reduces to the Bayes-Laplace prior.

Under the Jeffreys prior, it can be seen that
\begin{eqnarray*}
	 \lim\limits_{N\to\infty}P(R=N|T=n)=\frac{(n+1)k}{(n+1)k+1-2k},
\end{eqnarray*}
which for any large value of $n$ and $k\neq0$ is not zero. Specifically, for $k=\frac{1}{4}$, the above limit is $(n+1)/(n+3)$ which increases in $n$, and for large $n$ attains one as a limit.

Jeffreys' prior thus yields a result which accords well with the intuition of experimental scientists, but its construction, though ingenious, is ad hoc. An improvement over Jeffreys' prior, vis a vis a faster rate of convergence to one (as a function of n) as well as a justification that is grounded in information theoretic arguments, is a prior proposed by Bernardo (1985).

\subsection{Bernardo's Prior}

The essence of Bernardo's prior is a large point mass at $R=N$ only, as opposed to Jeffreys' large point masses at both $R=0$ and at $R=N$. Here
\begin{eqnarray*}
  P(R=N)= \bigg \{
  \begin{array}{l}
  \frac{1-k}{N}, \quad r=0,1,\cdots,N-1 \\
  k, \quad \quad r=0, N.
  \end{array}
\end{eqnarray*}
Under Bernardo's prior, it can be seen that
\begin{eqnarray*}
	\lim\limits_{N\to\infty}P(R=N|T=n)=\frac{(n+1)k}{(n+1)k+(1-k)},
\end{eqnarray*}
and this for $k=1/2$, yields $\lim\limits_{N\to\infty}P(R=N|T=n)=(n+1)/(n+2)$ which converges to one faster than Jeffreys' $(n+1)/(n+3)$ with its comparable $k$ of $1/2$, split between $R=0$ and $R=N$.

Bernardo's improvement over Jeffreys' prior via its grounding in information theoretic ideas, is a noteworthy development; see Section 3.2.

\subsection{A Portmanteau Prior}

The matter of \textit{\textbf{certifying}} a product, as a way of ensuring its future performance, is a topic in reliability and risk analysis that is akin to proving a law of nature. An archetypal example is a piece of pre-tested and debugged software which will now be tested $n$ times, under different inputs, to see if it satisfactorily processes them all. Once it does so successfully, the software is certified and released for actual use. In Singpurwalla and Wilson (2004) a prior distribution for R, which satisfies certain criteria germane to efficient certification testing is motivated and proposed. Whereas the details of this prior are cumbersome to present, Figure 1 below illustrates its general character.

There are finite point masses at $R=0$ and $R=N$, with $q\in(0,1)$, $k>0$, and $\lambda>0$, as parameters. The parameter $q$, together with a function $f(x)=kx^2$, for $0<x<\infty$, controls the rate of exponential decay (and revival) over $R=0,1,2,\cdots,N$ and $\lambda$ controls the sizes of the point masses at $R=0$ and $R=N$. This prior is not unlike Jeffreys' prior save for the exponential decay and revival between the point masses.\\
\begin{figure}[!ht]
  \centering
  \includegraphics[scale=0.5]{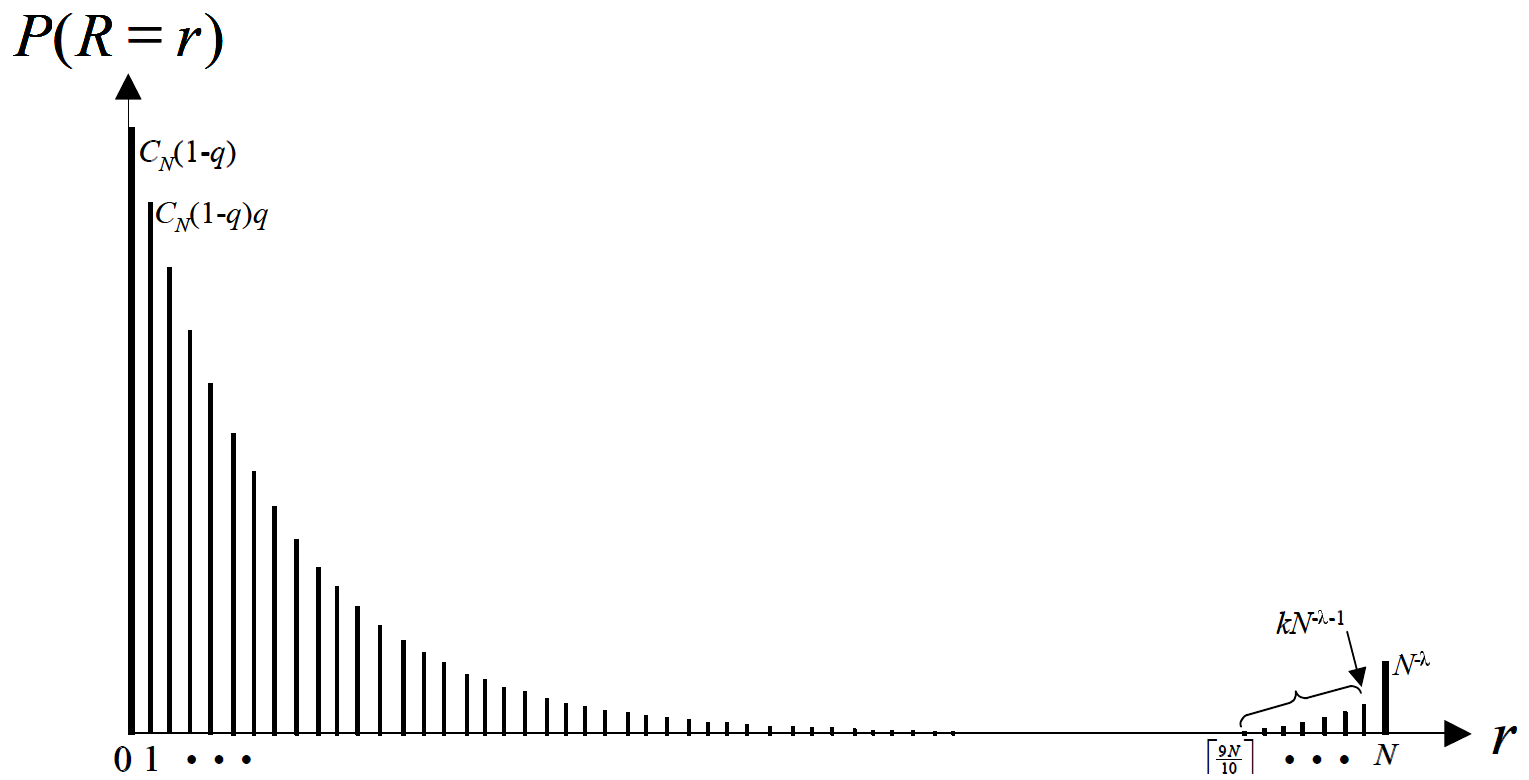}
  \caption{A Portmanteau Prior}
\end{figure}

For $k=5$, $\lambda=3$, and $q=0.5$, Figure 2 illustrates the behavior of $\lim\limits_{N\to\infty}P(R=N|T=n)$, as $n\to\infty$.
\newpage
\begin{figure}[!ht]
  \centering
  \includegraphics[scale=0.65]{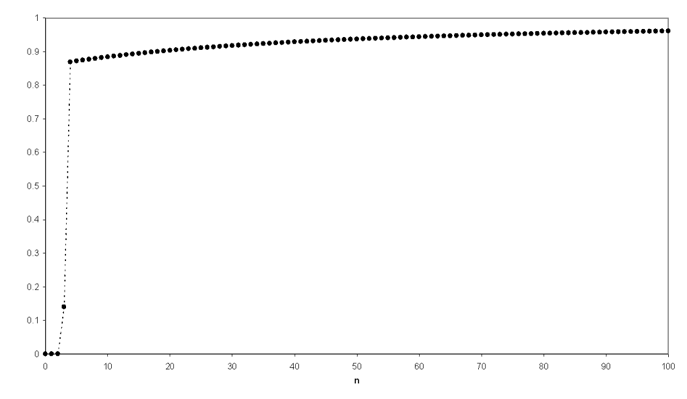}
  \caption{Probability of Non-Failure as $n$,$N\to\infty$}
\end{figure}
A virtue of this prior is that it leads to an assertion of an item's high probability to not fail only after a few successful trials, and this probability monotonically increases to one with additional successful tests. A prior such as this will be germane to testing for certification when the item in question has been previously vetted, and most of its flaws eliminated. Like the Jeffreys' and the Bernardo's prior, this prior also leads to results that accord with the intuition of scientists.

A noteworthy feature of the four classes of priors discussed above, is that there is a non-zero probability assigned to all the $N$ values that $R$ can take. In the contexts considered, an assignment like this is understandable.

\section{Propensities and Priors on Propensities}

We stated that the notion of ``almost identical trials'' is to be taken as an undefined primitive. In Section 1 and 2, we leaned on this primitive, and following the Bayes-Laplace prescription of assigning prior distributions on observable quantities, have proposed several classes of priors on $R$. Recall that $R$ being the unknown number of successes in $N$ trials is, in principle, observable, and thus any personal probability assignment on $R$ when viewed as a 2-sided bet can eventually be settled.

An approach alternate to that of Section 2, to be described in this section, has as its foundation de Finetti's (1974) notion of \textit{\textbf{exchangeability}}. This notion endeavors to make more precise ones sense of almost identical trials. Exchangeability is a judgment which implies \textit{\textbf{permutation invariance}}. When an infinite sequence of Bernoulli trials $X_1, X_2, \cdots,$ is declared exchangeable, de Finetti's famous representation theorem comes in play. It claims that for every $n\geq 1$, and every sub-set $X_1,\cdots,X_n$ of $X_1, X_2,\cdots, $ with $X_i=1(0)$ if the i-th trial is a success (failure):
\begin{eqnarray*}
  P(X_1=x_1,\cdots,X_n=x_n) &=& \int\limits_0^1 \prod\limits_{i=1}^n P(X_i=x_i|p)\Pi(p)dp \\
  &=& \int\limits_0^1 p^{\sum x_i}(1-p)^{n-\sum x_i}\Pi(p)dp
\end{eqnarray*}

Here $P$ and $\Pi$ denote personal probabilities of the $X_i$'s, $i=1,\cdots,n$, and a \textit{\textbf{fictional unknown quantity}} [cf. De Groot(1996), p.135], $p$, respectively; $0\leq p\leq1$. Statisticians refer to $p$ as a Bernoulli \textit{\textbf{parameter}}, whereas Good (1965) calls p a physical (or \textit{\textbf{objective probability}}); de Finetti refers to $p$ as a \textit{\textbf{chance}}. None of these labels provide any sense of what $p$ means. The quantity $\Pi(p)$ is ones personal prior probability distribution of $p$.

It is important to remark that under a personalistic theory of probability, $p$ should not be interpreted as a personal probability. Doing so would tantamount to looking at $\Pi(p)$ as a personal probability of a personal probability, and this cannot make $\Pi(p)$ operational [see Marschak (1975), p. 121-153 for a discussion]. We find it useful to look at $p$ as a propensity (see Section 3.1), in the sense of Popper (1969), and in a manner akin to Kolmogorov (1969).

In specifying $\Pi(p)$ one is assigning a personal probability to an ``unobservable fictional quantity'' $p$, and this would go against the Mach-Einstein dictum of \textit{operationalization}, because $\Pi(p)$ as a 2-sided bet on any value of $p\in [0,1]$ can never be settled. However an entity like $\Pi(p)$, as a mainstay of applied Bayesian statistics, is so commonly used that it warrants comment. Specifically $\Pi(p)$ enables one to automate an application of Bayes' Law in the light of new information, as an enforcer of coherence, and this in turn enables one to place bets on observables via their predictive probabilities.

\subsection{The Notion of Propensity}

The notion of a propensity was introduced by Karl Popper (1959) in connection with his attempts to interpret quantum theory. However, propensity is also implicit in the 1973 writings of de Finetti (1974), Kolmogorov (1969), and Kendall (1959). By propensity it is meant a physical tendency for the occurrence, or not, of a certain event, say success, under repeated incidences of its possible occurrence. An archetypal example is the occurrence of heads in infinite tosses of a coin under almost identical conditions. Since indefinitely tossing a coin under almost identical conditions is metaphysical, propensities are not observable, therefore not measurable, and thus not actionable. Indeed, propensities are not probabilities [cf. Humphreys (1985)], and certainty not personal probabilities. They encapsulate an interaction between the outcome of a random phenomenon and the conditions under which the phenomenon occurs [cf. Kolmogorov (1969)]. Observed relative frequencies, being the outcome of a finite number of occurrences, are not propensities either. Relative frequencies are viewed as manifestations of propensities. Propensities are unobservable abstractions useful for obtaining predictive distributions via Bayes' Law, once they are endowed with personal probabilities.

To the best of our knowledge, the term ``propensity'' has appeared in at least three different contexts. The first is in the context of quantum theory mentioned above; the second in educational testing and the social sciences wherein ``propensity scores'' are used for assigning treatment effects. A propensity score is de facto, a conditional probability, and thus has no bearing on what we consider here. The third context in which the term propensity has appeared is that of the chemical Langevin equation [cf. Gillespie (2000)], and comes closest in spirit to what we think of as here.

In the context of exchangeable Bernoulli trials, an extension of de Finetti's theorem shows the propensity $p$ as $\lim\limits_{n\to\infty}\frac{1}{n}\sum\limits_1^n X_i$. Since this limit cannot be known de Finetti endows $p$ with a prior distribution $\Pi(p)$, and in so doing provides a foundation for Bayesian inference under Bernoulli trials. In what follows we first discuss a commonly used choice for $\Pi(p)$, and then introduce a new choice which is a scale transform of the commonly used choice. This new choice could be more suitable for the contexts considered in this paper.

\subsection{The Beta Family of Priors on Propensity}

Since $0\leq p\leq 1$, a natural choice for $\Pi(p)$ would be a standard beta distribution with constants (parameters) $\alpha >0$ and $\beta >0$. Different choice for $\alpha$ and $\beta$ enable one to represent different shapes for $\Pi(p)$, and thus different judgments about the propensity $p$. For example, $\alpha=\beta=1$ makes $\Pi(p)$ uniform over $p$, so that $\Pi(p)$ can be seen as an analogue of the Bayes-Laplace public prior on $R$. Identical values of $\alpha>1$, $\beta>1$ make $\Pi(p)$ symmetric around $p=1/2$, whereas unequal values of $\alpha<(>)\beta$ make $\Pi(p)$ skewed to the right(left). To encapsulate the prior opinion that values of $p$ are closer to one than to zero, one would set $\alpha > \beta$. Similarly with  $\alpha < \beta$.

$\Pi(p)$ will be L-shaped, when $\alpha<1$ and $\beta>1$; it is J-shaped when $\alpha>1$ and $\beta<1$, and U-shaped when $\alpha<1$ and $\beta<1$ [cf. Singpurwalla(2006), p.130]. In all the above cases, there is no probability mass at $p=0$ and/or $p=1$. However, when $\alpha <1 $ and $\beta =1 $, $\Pi(p)$ is L-shaped with a probability mass at $p=1$; with $\alpha=1$ and $\beta<1$, $\Pi(p)$ is J-shaped with probability mass at $0$; see Figure 3 and 4.

Of the above choices, the ones which appear to be the closest in terms of relevance for the scenarios of asserting a law, or certifying a product, are the U-shaped $\Pi(p)$, with $\alpha<1$ and $\beta<1$, the L-shaped prior with $\beta=1$ and $\alpha<1$ or the J-shaped prior with $\alpha=1$ and $\beta<1$. Of these the L-shaped prior with a finite positive probability at $p=1$ seems most promising. All of the above priors cover the entire $[0,1]$ range of $p$, and there could be scenarios wherein subjective opinion could be such that one need focus only on a segment of this range, say the range $[\omega,1]$ for some $\omega >0$. This matter is taken up later in Section 3.4.

For now, it is well known that when $\Pi(p)$ ha a beta distribution with parameters $\alpha$, $\beta>1$, then having observed $n$ successes in $n$ Bernoulli trials the predictive probability that $N$ future trials will yield $N$ successes is given by \textbf{beta-binomial} [cf. Singpurwalla(2006), p.158] as
\begin{eqnarray*}
	\frac{B(\alpha+n,\beta)}{B(\alpha+n+N,\beta)}, \quad where \ B(a,b)=\frac{\Gamma(a+b)}{\Gamma(a)\Gamma(b)}.
\end{eqnarray*}
It can be seen that for any fixed $n$, no matter how large, the limit of the above probability, as $N\to\infty$, is zero. This conclusion is analogous to that of Bayes and Laplace when a public prior is assigned to $R$; it too will not accord with the intuition of experimenters.

Alternatively, suppose that the prior $\Pi(p)$ is L-shaped with a probability mass at $p=1$; this is obtained via a beta distribution with parameters $\alpha<1$, and $\beta=1$. Specifically, for $0\leq p\leq 1$,
\begin{eqnarray*}
	\Pi(p)=\frac{\Gamma(\alpha+1)}{\Gamma(\alpha+1)}p^{\alpha-1}= \frac{\alpha \Gamma(\alpha)}{\Gamma(\alpha+1)}\frac{1}{p^{1-\alpha}}=\frac{\alpha}{p^{1-\alpha}}.
\end{eqnarray*}	

As can be seen from Figure 3, this prior emphasizes small values of $p$, even though there is no probability mass at $p=0$.
\newpage
\begin{figure}[!ht]
  \centering
  \includegraphics[scale=0.5]{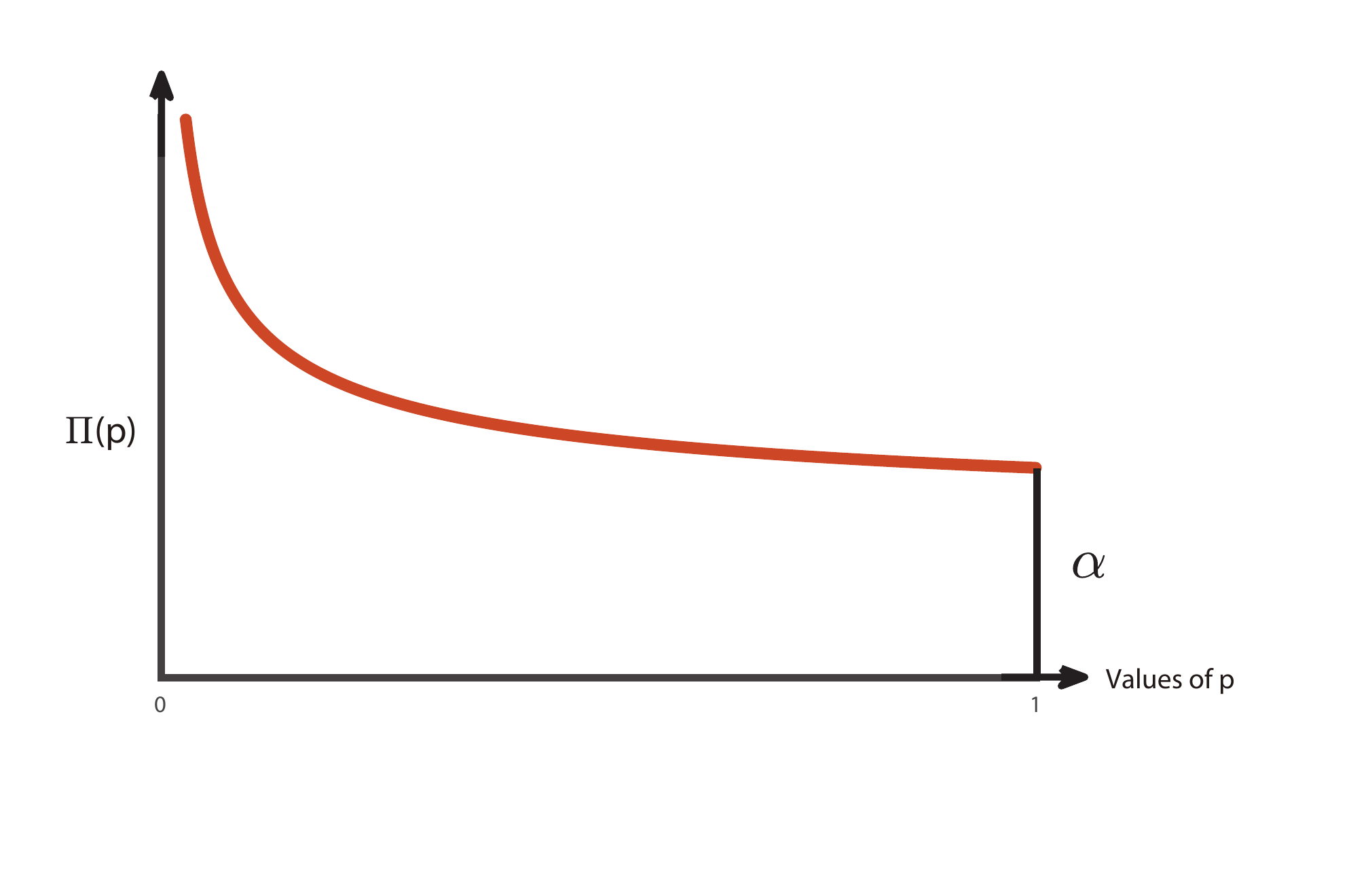}
  \caption{L-Shaped Prior for Propensity $p$ ($\alpha<1$, $\beta=1$)}
\end{figure}

Under the above prior, if $n\geq 1$ Bernoulli trials yield all $n$ successes, then via routine calculations it can be seen that the posterior distribution of $p$ is of the form $(n+\alpha)p^{n+\alpha-1}$ which is a beta distribution with parameters $(n+\alpha)$ and $1$. The effect of testing is to increase the probability of $p$, at $p=1$, from its original $\alpha$, to $(\alpha+n)$. A consequence of this posterior of $p$, is that the predictive probability of observing all $N$ successes in $N$ future trials is given by the beta-binomial distribution as $(\alpha+n)/(\alpha+n+N)$, which as $N\to\infty$ goes to zero; see the Appendix for details. Here again, this result would be against the grain of experimentalists, making a prior such as this unsuitable for certification and validation..

What if the prior chosen is J-shaped with a finite probability at $p=0$, and an infinite probability density at $p=1$?; see Figure 4. This prior, which is similar in spirit to that of Bernardo's, can be obtained via a beta distribution with parameters $\alpha=1$ and $\beta<1$; that is, $\Pi(p)=\beta/(1-p)^{(1-\alpha)}$.
\newpage
\begin{figure}[!ht]
  \centering
  \includegraphics[scale=0.5]{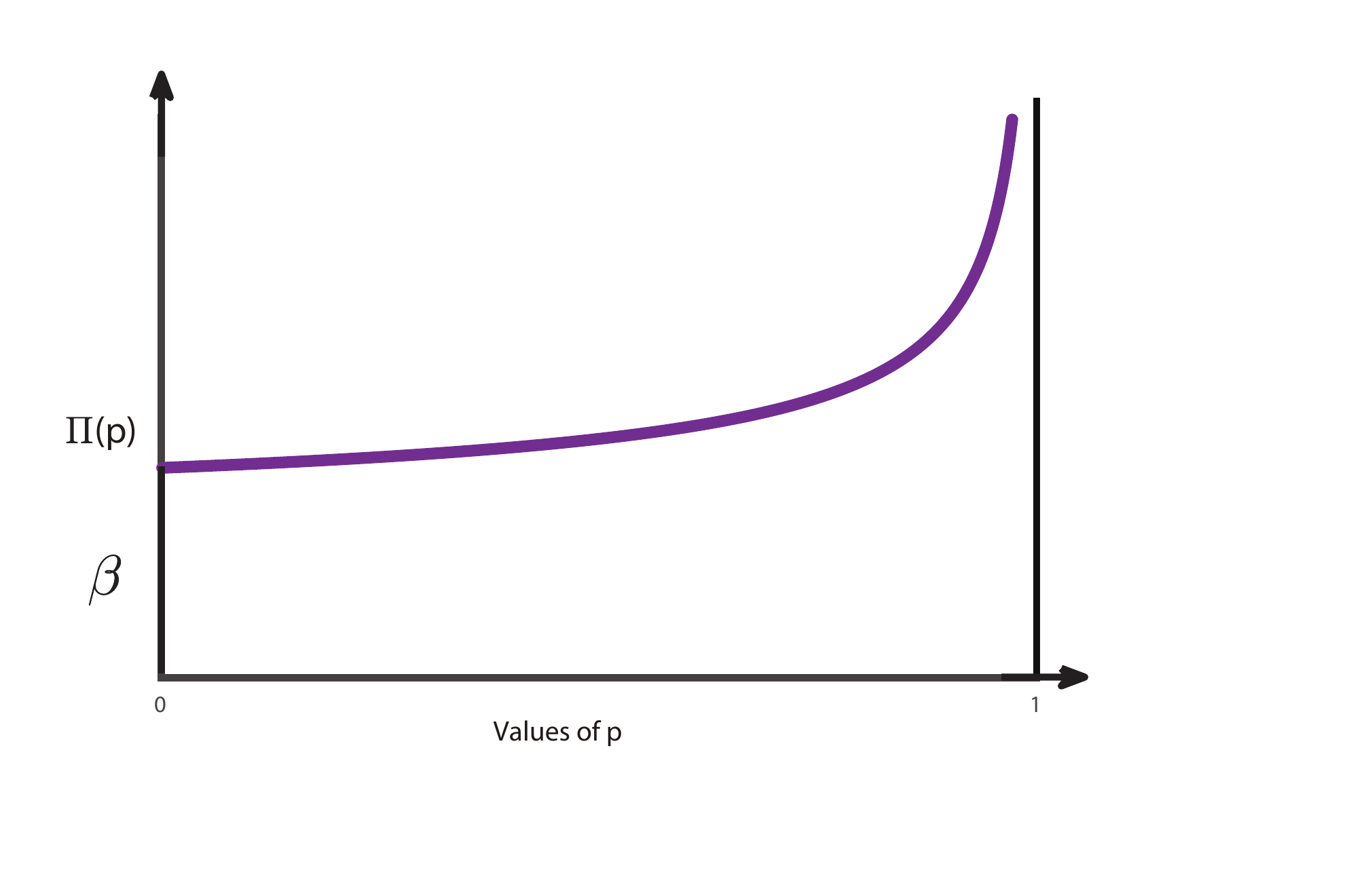}
  \caption{J-Shaped Prior for Propensity $p$ ($\alpha =1$, $\beta <1$)}
\end{figure}

Under the above prior, if $n\geq1$ Bernoulli trials yield all $n$ successes, then the posterior distribution of $p$ is, for some constant $C$, of the form $C\beta p^n(1-p)^{\beta-1}$; this is a beta distribution with parameters $(n+1)$ and $\beta$. Consequently, the predictive probability of observing $N$ successes in $N$ future trials is proportional to $B(n+1,\beta)/B(n+1+N,\beta)$. Because $0<\beta<1$, the limit of this probability, as $N\to\infty$, is for small to moderate $n$ zero, whereas it approaches one as $n$ becomes large. As is shown in the Appendix, the predictive probability is the product of $N$ terms, each of the form $(n+N)/(n+N+\beta)$, which for any $\beta\in(0,1)$ is a number less than one when $n$ is small or moderate, and is approximately equal to one when both $n$ and $N$ are large. Figure 5 illustrates the behavior of the predictive probability for $\beta=0.1$ and $N=10,000$, over a range of values of $n$. This behavior of predictive probability better encapuslates the intuition of experimentalists than the other predictive probabilities discussed hereto, because ones certitude about a law should increase as the number of successful test cases increase.

Thus, like the priors of Jeffreys, Bernardo, and the portmanteau, of Sections 2.2, 2.3, and 2.4, a prior such as this is viable for certification and validation, provided that $n$ is large.
\begin{figure}[!ht]
  \centering
  \includegraphics[scale=0.5]{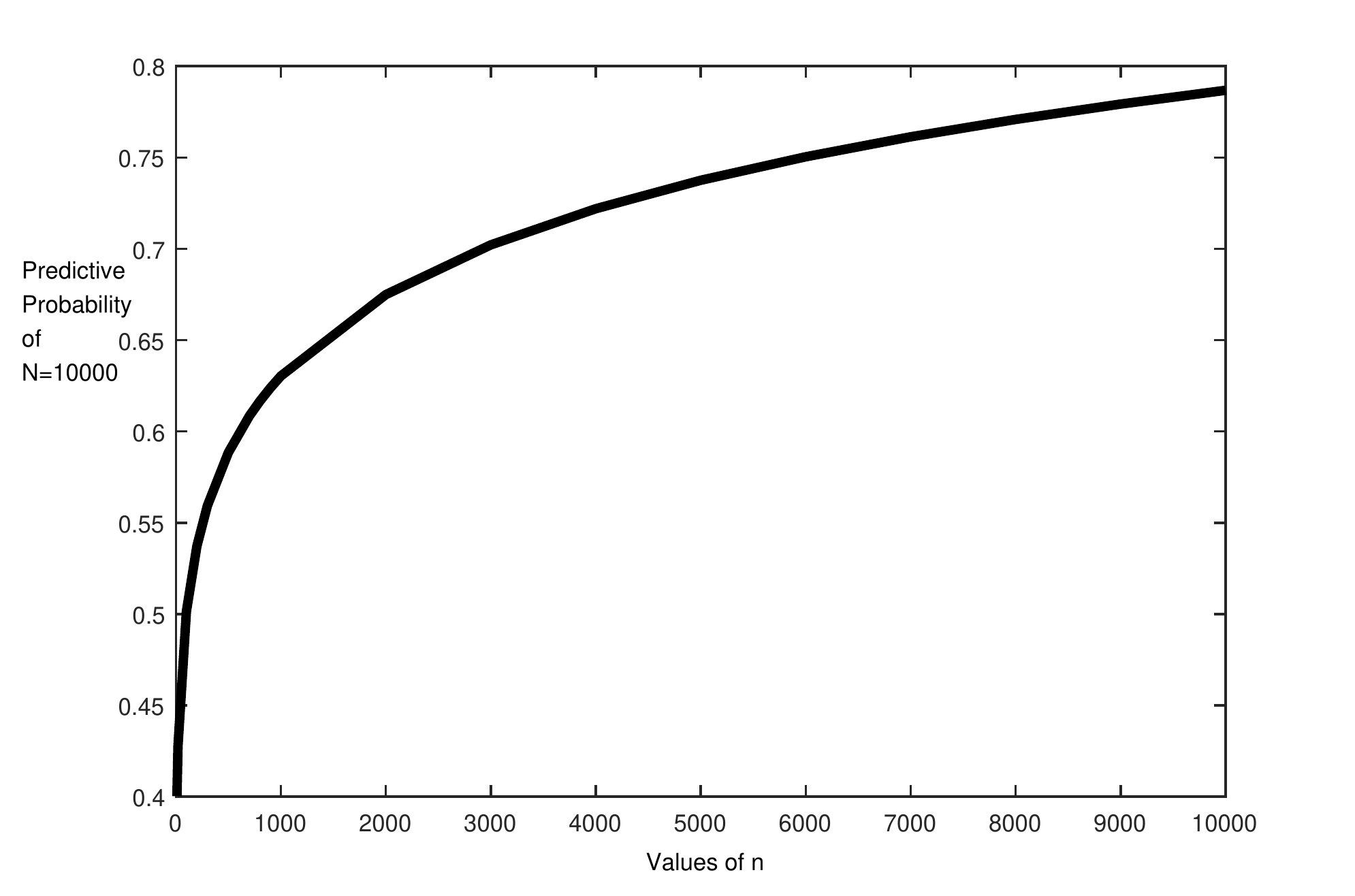}
  \caption{Predictive Probability Under a J-shaped Prior With $\beta=0.1$}
\end{figure}

\subsection{A Reflected Scale Transformed Beta Family}

The beta family of prior distributions considered in Section 3.2 covers the entire range of $0$ to $1$ for values of the propensity $p$. In many applications, especially those pertaining to system certification, this could be seen as being restrictive because the performance of systems improves over time as a consequence of the process of debugging and successive fixes. In such circumstances, a prior on $p$ over the range $[\omega,1]$ for some value of $\omega>0$, may be judged more appropriate. One way to achieve this form of left truncation is via a transformation of the beta distribution of $p$, by scaling it by the factor $e^{-\eta}$, for some $\eta\geq0$, and then taking its reflection. Specifically, we let $p=1-e^{-\eta}z$, where $\eta\geq0$, with $z$ having a beta distribution with parameter $\alpha>0$ and $\beta>0$. When $\eta=0$, the distribution of $p$ will be a standard beta over $[0,1]$, and for values of $\eta>0$, the range of values of $p$ will be restricted to the left, so that the distribution of $p$ will concentrate its mass more and more towards one. See Figure 6, wherein $\alpha=3$, $\beta=2$, and $\eta=0, \ 0.5,$ and $1$, respectively; the support of the distribution of $p$ decreases with increasing values of $\eta$.
\begin{figure}[!ht]
  \centering
  \includegraphics[scale=0.5]{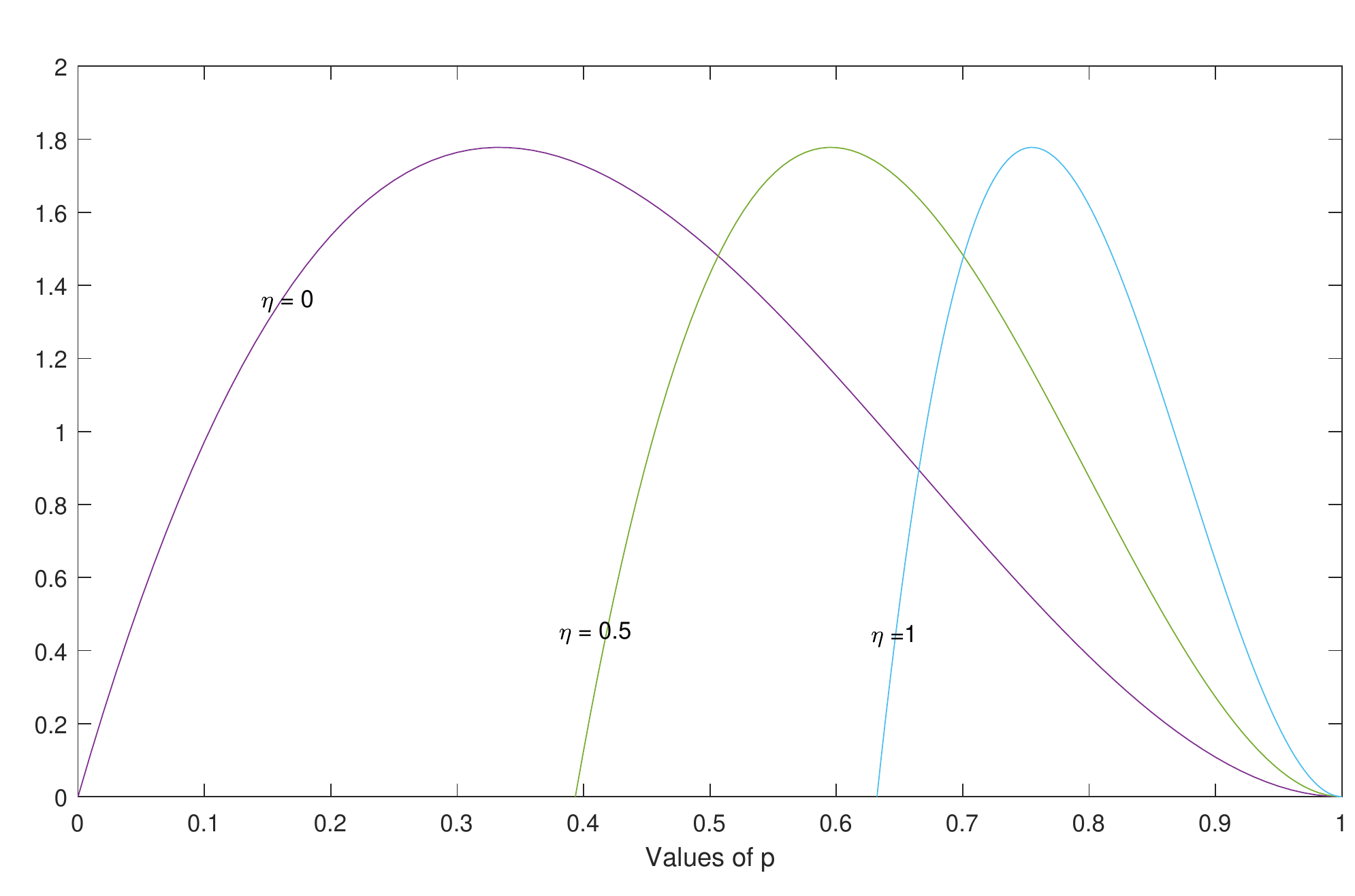}
  \caption{Plots of Left Truncated Beta Distribution of $p$.}
\end{figure}

Verify that for $c=e^{-\eta}$, with $\eta>0$, the prior distribution of $p$ is,\\[-10pt]
\begin{eqnarray*}
	\Pi(p)=\frac{\Gamma(\alpha+\beta)}{\Gamma(\alpha)\Gamma(\beta)}(\frac{1}{c})^{\alpha+\beta-1}(1-p)^{\alpha-1}(c-1+p)^{\beta-1}, \quad 1-c\leq p\leq 1;
\end{eqnarray*}  \\[-10pt]
this is a \textit{\textbf{shifted beta}} distribution with parameters $\alpha$ and $\beta$, over $(1-c)$ and $1$. Because this distribution does not have a probability mass at $p=1$, it will be unsuitable for the purpose of certification and validation. The distribution is meritorious all the same because it has been used by us as a model for the posterior distribution of the threshold $1-c$; see Section 3.4.1.

Were one to set $\alpha=1$ and $\beta<1$, so that the distribution is better aligned with the J-shaped distribution of Figure 4, the prior distribution of $p$ would simplify as \begin{eqnarray*}
	\Pi(p)=\beta(\frac{1}{c})^\beta (c-1+p)^{\beta-1}, \quad (1-c)\leq p\leq1;
\end{eqnarray*}
see Figure 7.
\newpage
\begin{figure}[!ht]
  \centering
  \includegraphics[scale=0.6]{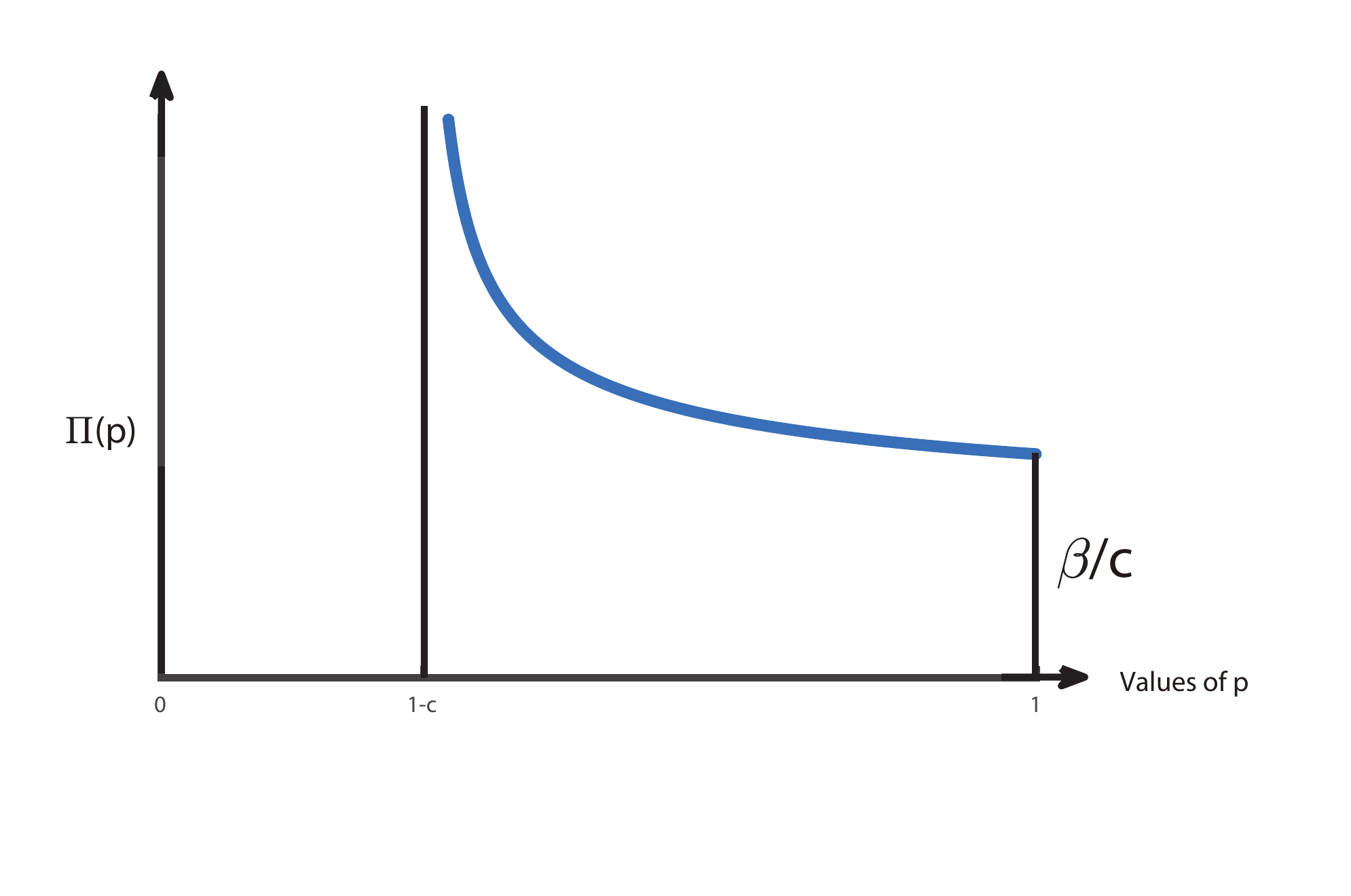}
  \caption{L-Shaped Prior by Transformation and Reflection ($\alpha=1$, $\beta<1$, and $c=e^{-\eta}$)}
\end{figure}

Whereas this prior, which is analogous to the L-shaped prior of Figure 3, has a probability mass of $\beta/c$ at p=1, it will also have an infinite probability density at $p=1-c$, making it unsuitable for use in certification and validation. Indeed with $n$ successes out of $n$ trials, the posterior distribution of $p$ will continue to be a beta distribution with parameters $(n+1)$ and $\beta$ over $(1-c)$ and $1$. This distribution will have a probability mass of
\begin{eqnarray*}
\frac{\Gamma(n+1,\beta)}{\Gamma(n+1)\Gamma(\beta)}c^{\beta-1}, \quad at \ p=1,
\end{eqnarray*}
which, even for large $n$, is only slightly greater than $\beta/c$, for $\beta<1$, suggesting an insensitivity of the distribution at $p=1$ to the observed success data.

Based on the above, as well as the illustrations of Figure 6, it appears that reflections of scale transformed beta distributions are inappropriate for the task of certification and validation. We therefore consider in section 3.4 transformations entailing location shifts of suitable beta distributions. The illustrations of Figure 6 and 7 are presented here for the sake of completeness, though Figure 6 is germane to the material of Section 3.4.1.

\subsection{Left Truncated Beta Family}

Suppose that the prior distribution of $p$ is a location shifted beta distribution with parameters $\alpha=1$ and $\beta<1$, over the range $[\omega,1]$, where $0<\omega \leq 1$ is a specified parameter. The parameter $\omega$ is analogous to the parameter $(1-c)$ of Figure 7. Verify that the probability density of $p$ is of the form\\ [-10pt]
\begin{eqnarray*}
	\Pi(p)=\frac{\beta}{(1-\omega)^\beta}(1-p)^{\beta-1}, \quad for \ \omega\leq p \leq 1.
\end{eqnarray*}
This distribution places a probability mass of $\beta/(1-\omega)$ at $p=\omega$, and an infinite probability density at $p=1$; see Figure 8.
\begin{figure}[!ht]
  \centering
  \includegraphics[scale=0.6]{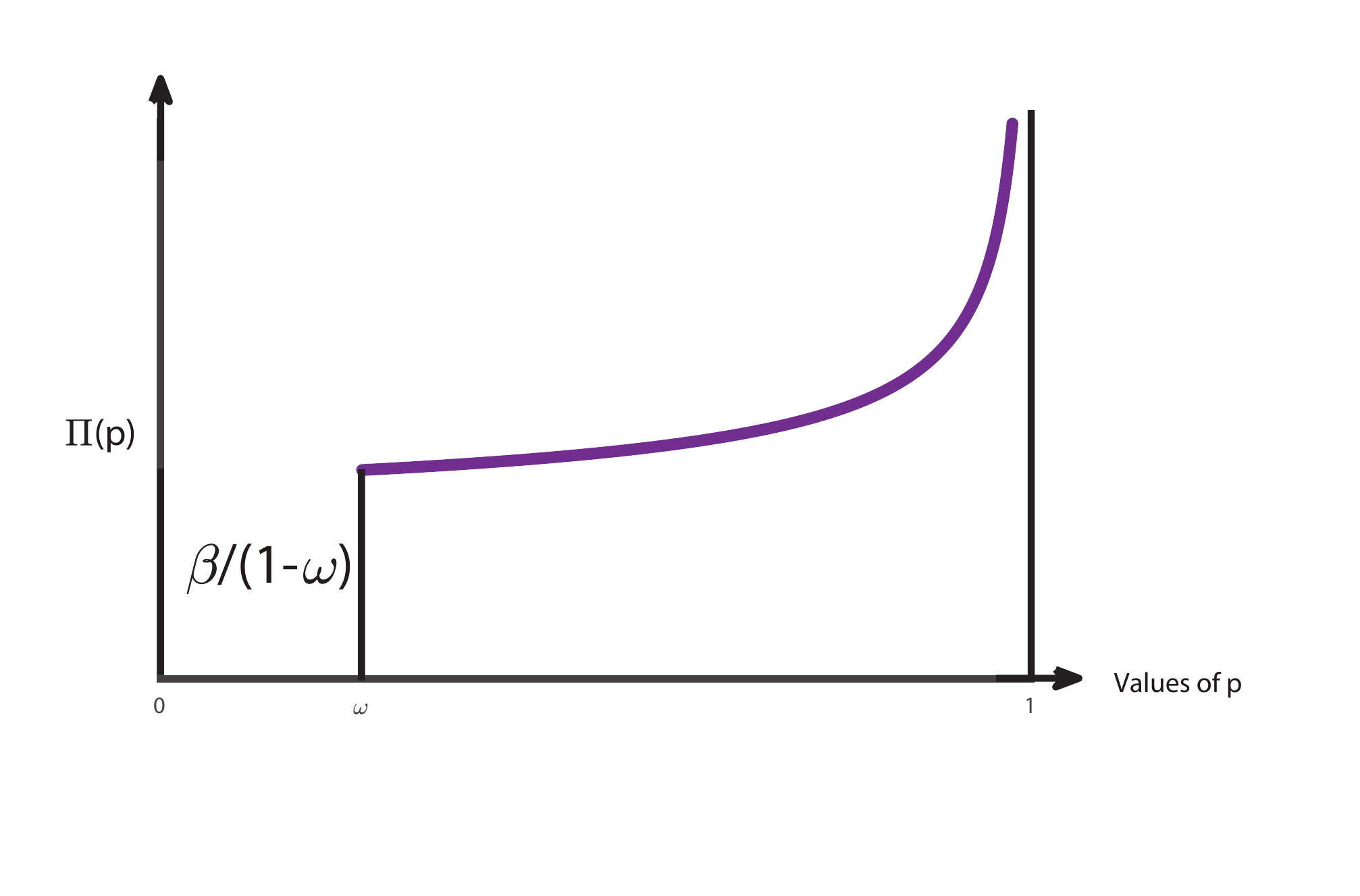}
  \caption{Location Shifted Beta ($\alpha=1$, $\beta<1$, and $\omega>0$)}
\end{figure}

If all $n$ out of $n$ Bernoulli trials result in a success, then the posterior distribution of $p$ is also a beta distribution with parameters $(n+1)$ and $\beta$, over the range $[\omega,1]$. Specifically, with $\omega$ specified,
\begin{eqnarray*}
	\Pi(p;n)=\frac{\Gamma(n+\beta+1)}{\Gamma(n+1)\Gamma(\beta)}\frac{(p-\omega)^n(1-p)^{\beta-1}}{(1-\omega)^(n+\beta)}, \quad \omega\leq p \leq 1.
\end{eqnarray*}
This posterior distribution has an infinite probability density at $p=1$, but here the (prior) probability mass of $\beta/(1-\omega)$ at $p=\omega$, vanishes. This latter feature is to be expected, since none of the $n$ Bernoulli trials have experienced a failure.

It is easy to verify that the predictive probability of observing all $N$ successes in $N$ future trials is given as $B(n+1,\beta)/B(n+N+1,\beta)(1-\omega)^{n+\beta}$, which when n is moderate to large, and $\omega,\beta<1$, converges to one as $N\to\infty$. This result is analogous to that associated with the prior Figure 4, save for the extra $(1-\omega)^{n+\beta}$ term in the denominator. The effect of this extra denominator term is to enable a faster rate of convergence to one as compared to that given by $B(n+1,\beta)/B(n+N+1,\beta)$ alone.

The location-shifted prior of this section is therefore suitable for asserting certification and validation, and more so than that associated with the prior of Figure 4, because of a faster rate of convergence. Provided that one is able to specify $\omega$ in a meaningful fashion, the prior of Figure 7 seems to be the most rewarding of the priors on propensity that we have considered thus far. However, specifying a precise value for $\omega$ would be enigmatic. A challenge therefore is how best to specify a prior on $\omega$, and based on $n$ out of $n$ successful Bernoulli trials, how to induce its posterior. This matter is taken up below.

We start by recalling that given $\omega$, $\omega\in[0,1]$, the posterior distribution of $p$, having observed $n$ out of $n$ is:\\
\begin{eqnarray*}
\Pi(p|\omega;(n \ out \ of \ n \ successes))=\frac{\Gamma(n+\beta+1)}{\Gamma(n+1)\Gamma(\beta)}\frac{(p-\omega)^n(1-p)^{\beta-1}}{(1-\omega)^(n+\beta)}, \quad \omega\leq p \leq 1.
\end{eqnarray*}

Were we to average out this posterior distribution with respect to a posterior distribution of $\omega$ (in the light of $n$ out of $n$ successes) we would obtain the posterior of $p$ unconditional on $\omega$. However, obtaining a posterior of $\omega$, via the usual Bayesian prior to posterior iteration, poses a difficulty when one does not have a probability model for inducing a likelihood function for $\omega$. All the same, there is nothing within the Bayesian paradigm which mandates that the only way to obtain posterior distributions is via a legislated application of Bayes' Law. Indeed, the philosopher Richard Jeffrey has proposed an alternative to Bayes' Law [cf. Shafer (1981), Diaconis and Zabell (1982)], called \textbf{\textit{Jeffrey's Rule of Conditioning}}. Furthermore, since the prior is to be subjectively specified, and so can the likelihood be [were one not to subscribe to the philosophical \textbf{\textit{principle of conditionalization}}, cf. Williams (1980)], one is also at liberty to subjectively specify a posterior distribution, subject to the usual constraints of coherence. This is what we choose to do, as the material below indicates a possible approach.

\subsubsection{Subjectively Specified Posterior of Location}

To facilitate a subjective specification of the posterior distribution of $\omega$ which is coherent, we lean on the plots of Figure 6, each appropriate to a value of n successes (out of $n$ trials). These plots encapsulate the feature that their thresholds move towards one, as $n$ increases. The functional form describing these plots is taken to be the posterior distribution of $\omega$, which is a shifted beta distribution over $(1-c)$ and $1$, with parameters $a$ and $b$. Specifically, we suppose that
\begin{eqnarray*}
\Pi(\omega; a, b, c)=\frac{\Gamma(a+b)}{\Gamma(a)\Gamma(b)}(\frac{1}{c})^{a+b-1}(1-\omega)^{a-1}(c-1+\omega)^{b-1}, \quad 1-c\leq \omega \leq 1.
\end{eqnarray*}

In table 1, we indicate possible choices for $a,b$ and $c$ for different values of $n$, assigning all $n$ successes. The entries of the Table 1 are by no means unique.
\newpage

\begin{table}[htbp]
\centering
\begin{tabular}{|p{50pt}<{\centering}|p{50pt}<{\centering}|p{50pt}<{\centering}|p{50pt}<{\centering}|}
\hline
m   & a & b & c      \\
\hline
1   & 1 & 1 & 1     \\
2   & 2 & 2 & 1     \\
5   & 2 & 3 & 1      \\
10  & 3 & 4 & 0.95   \\
25  & 3 & 4 & 0.9    \\
50  & 3 & 4 & 0.75   \\
75  & 4 & 5 & 0.5    \\
100 & 4 & 5 & 0.25   \\
1000 & 4 & 5 & 0.05  \\
\hline
\end{tabular}
\caption{Poster Distribution of $\omega$ as a Function of $n$}
\end{table}

Averaging out $\omega$ with respect to the above posterior distribution of $\omega$, in $\Pi(p|\omega;n\dot)$ givn before, gives us the posterior distribution of $p$, unconditional on $\omega$ as:
\begin{eqnarray*}
\Pi(p;n,\dot)=\frac{\Gamma(n+\beta+1)}{\Gamma(n+1)\Gamma(\beta)}\frac{\Gamma(a+b)}{\Gamma(a)\Gamma(b)}(\frac{1}{c})^{a+b-1}(1-p)^{\beta-1}\int\limits_{1-c}^1\frac{(p-\omega)^n(c-1+\omega)^{b-1}}{(1-\omega)^{n+\beta+1-a}}d\omega,
\end{eqnarray*}
an expression difficult to evaluate in closed form. Its numerical evaluation, is straightforward.

Whereas knowing the $\omega$-averaged posterior distribution of $p$ is of limited interest, the more relevant entity to know is the predictive probability of $N$ out of $N$ successes in $N$ future trials. With $\omega$ specified, this probability is given as $B(n+1,\beta)/B(n+N+1,\beta)(1-\omega)^{n+\beta}$. Averaging out $\omega$ with respect to its posterior distribution, gives the unconditional predictive probability as:
\begin{eqnarray*}
\frac{B(n+1,\beta)}{B(n+N+1,\beta)}\frac{\Gamma(a+b)}{\Gamma(a)\Gamma(b)}(\frac{1}{c})^{a+b-1}\int\limits_{1-c}^1\frac{(c-1+\omega)^{b-1}}{(1-\omega)^{n+\beta+1-a}}d\omega.
\end{eqnarray*}
For $n$ large, the integral term simplifies as $\int\limits_{1-c}^1\frac{(c-1+\omega)^{b-1}}{(1-\omega)^n}d\omega$, and this can be numerically assessed.

\section{Summary and Conclusions}

The aim of this paper is the articulate the conditions and circumstances in which observing a slew of successes, and no failures, enables one to assert a law of nature, or to claim the certification of an entity. These circumstances pertain to the characteristics of ones prior opinion, assuming that it is honest and genuine, and how it is expressed. One cannot assert certitude if prior opinion is of the ``objective'' type, or is encapsulated via distributions for which a predictive probability converges to zero. There do exist priors for which the predictive probability goes to one, for moderate, or even small values of the number of items tested. If such priors reflect true opinion, then their use could be beneficial, costwise. In an adversarial circumstance, a use of such priors can be subject to a challenge.

It is not the purpose, nor the intent, of this paper to show, or to make the case, that by manipulating priors, one can get what one wants to obtain certitude and claim certification. Rather, it is to make the case that by choosing certain priors one can be assured of getting what one is entitled to get, because in observing a large sequence of successes and no failures, a user of certain kinds of priors could face the dilemma of a \textbf{\textit{reductio ad absurdum}}.
\newpage

\section*{\large \centering Appendix}
\begin{center}
(Convergence of Posterior Predictive Probabilities)
\end{center}
{\normalsize
\begin{enumerate}[1.]
\item For the L-shaped prior, the predictive probability that all $N$ future trials will lead to success is
\begin{eqnarray*}
B(\alpha+n,1)/B(\alpha+n+N,1)  & =&  \frac{\Gamma(\alpha+n+1)}{\Gamma(\alpha+n)}\cdot \frac{\Gamma(\alpha+n+N)}{\Gamma(\alpha+n+N+1)} \\
 &=&  \frac{(\alpha+n)!}{(\alpha+n-1)!}\cdot\frac{(\alpha+n+N-1)!}{(\alpha+n+N)!} \\
  &=&  \frac{(\alpha+n)(\alpha+n-1)!}{(\alpha+n-1)!}\cdot\frac{(\alpha+n+N-1)!}{(\alpha+n+N)(\alpha+n+N-1)!}  \\
   &=&  \frac{\alpha+n}{\alpha+n+N} \to 0 \quad as \quad N\to \infty
\end{eqnarray*}
\item For the J-shaped prior, the predictive probability that all $N$ future trials will lead to success is proportional to $ B(n+1,\beta)/B(n+N+1,\beta) $
\begin{eqnarray*}
& =&  \frac{\Gamma(n+1+\beta)}{\Gamma(n+1)\Gamma(\beta)}\cdot  \frac{\Gamma(n+N+1)\Gamma(\beta)}{\Gamma(n+N+1+\beta)} \\
 &=&  \frac{(n+\beta)!}{n!}\cdot\frac{(n+N)!}{(n+N+\beta)!} \\
  &=&  \frac{(n+N)!}{n!}\cdot\frac{(n+\beta)!}{(n+N+\beta)!} \\
   &=&  \frac{(n+N)(n+N-1)\cdots(n+N-N)!}{n!}\cdot\frac{(n+\beta)!}{(n+N+\beta)(n+N+\beta-1)\cdots(n+N+\beta-N)!} \\
  &=&  \frac{(n+N)(n+N-1)\cdots(n+N-N+1)}{(n+N+\beta)(n+N+\beta-1)\cdots(n+N+\beta-N+1)}
\end{eqnarray*}
which for any $\beta<1$ is a product of $N$ terms, each close to one when $n$ gets large. The product tends to $1$ as $N\to\infty$.
\end{enumerate}}

\newpage
\section*{\large Acknowledgements}

The idea of scale transforming a beta distributed variable to induce a threshold was suggested to the first author by Professor N. Balakrishnan of Mc Master University. Whereas this idea did not fly in its intended spirit, it proved valuable for the subjective specification of posterior distributions. We thank Prof. Balakrishnan for this suggestion. The work reported here was supported by a grant from the City University of Hong Kong, Project Number 9380068, and by the Research Grants Council Theme-Based Research Scheme Grant T32-102/14N and T32-101/15R.

\newpage
\section*{\large References}
\quad \\ [-100pt]


\begin{thebibliography}{4}

\small

\bibitem{Bernardo1985}
{\sc Bernardo, J. M. (1985)}.
\newblock``On a Famous Problem of Induction.''
\newblock Trabajos de Estadistica, 36, 1:24-30.

\bibitem{De1974}
{\sc De Finetti, B. (1974)}.
\newblock``Theory of Probability: A Critical Introductory Treatment.''
\newblock Vol. 2, London, New York, Wiley. .

\bibitem{De1966}
{\sc De Groot, M. (1966)}.
\newblock``In Personal Probabilities of Probabilities.''
\newblock Theory and Decision (1975). 6 (2), p. 135.

\bibitem{Dia1982}
{\sc Diaconis, Persi,} and {\sc Zabell, Sandy. L. (1982)}.
\newblock``Updating Subjective Probability.''
\newblock Journal of the American Statistical Association, 77 (380), 822-830.


\bibitem{Gil2000}
{\sc Gillespie, D. T. (2000)}.
\newblock``The chemical Langevin equation.''
\newblock The Journal of Chemical Physics, 113(1), 297-306.

\bibitem{Good1965}
{\sc Good, I. J. (1965)}.
\newblock``The Estimation of Probabilities: An Essay on Modern Bayesian Methods.''
\newblock M.I.T. Press, Cambridge, MA.

\bibitem{Hum1985}
{\sc Humphreys, P (1985)}.
\newblock``Why Propensities Cannot be Probabilities.''
\newblock Philos. Review. 94, 557-570.

\bibitem{Jeff1961}
{\sc Jeffreys, H. (1961)}.
\newblock``Theory of Probability.''
\newblock Third Edition. Clarendon Press, Oxford, England.

\bibitem{Ke2015}
{\sc Ke, Q., Ferrara, E., Radicchi, F., and Flammini, A. (2015)}.
\newblock``Defining and identifying sleeping beauties in science.''
\newblock  Proceedings of the National Academy of Sciences, Vol. 112, No.24, pp 7426-7431.


\bibitem{Ken1959}
{\sc Kendall, M. G. (1959)}.
\newblock``On the Reconciliation of Theories of Probability.''
\newblock Biometrika, 36. 101-116.

\bibitem{Kol1969}
{\sc Kolmogorov, A. N. (1969)}.
\newblock``The Theory of Probability.''
\newblock In Mathematics: Its Content, Methods, and Meaning, Vol. 2, Part 3. A.D. Aleksandrov, A. N. Kolmogorov, and M. A. Lavrentav. Eds. M.I.T. Press, Cambridge, MA.

\bibitem{Mar1966}
{\sc Marschak, J. (1966)}.
\newblock``Do Personal Probabilities of Probabilities Have an Operational Meaning? In Personal Probabilities of Probabilities.''
\newblock Theory and Decision, 6(2), 127-133.

\bibitem{Pear1920}
{\sc Pearson, K. (1920)}.
\newblock``The Fundamental Problem of Practical Statistics.''
\newblock Biometrika, 13(1), 1-16.

\bibitem{Popper1959}
{\sc Popper, K. R. (1959)}.
\newblock``The Propensity Interpretation of Probability.''
\newblock The British Journal for the Philosophy of Science, 10(37), 25-42.

\bibitem{Shafer1981}
{\sc Shafer, G. (1981)}.
\newblock``Jeffrey’s Rule of Conditioning.''
\newblock Philosophy of Science, Vol. 48, No. 3, 337-362.

\bibitem{Sin2004}
{\sc Singpurwalla, N. D.,} and {\sc Wilson, P. (2004)}.
\newblock``When can Finite Testing Ensure Infinite Trustworthiness?''
\newblock J. Iranian Statist. Soc, 3, 1-37.

\bibitem{Sing2006}
{\sc Singpurwalla, N. D. (2006)}.
\newblock``Reliability and Risk: a Bayesian Perspective.''
\newblock John Wiley \& Sons, New York.




\bibitem{William1980}
{\sc Williams, R. (1980)}.
\newblock ``Bayesian Conditionalization and the Principle of Minimum Information.''
\newblock Brit. J. Phil Sci. 3x, 131-144.

\bibitem{Wilson1927}
{\sc Wilson, E.B. (1927)}.
\newblock ``Probable Inference, The Law of Succession, and Statistical Inference.''
\newblock Journal of the American Statistical Association. Vol. 22, No.158, pp 209-212.



\bibitem{Zabell1989}
{\sc Zabell, S. (1989)}.
\newblock ``The Rule of Succession.''
\newblock Erkenntnis 31; 283-321.



\end{thebibliography}
\end{document}